\pdfoutput=1

\documentclass{jpp}

\usepackage{graphicx}
\usepackage{natbib}

\ifCUPmtlplainloaded \else
  \checkfont{eurm10}
  \iffontfound
    \IfFileExists{upmath.sty}
      {\typeout{^^JFound AMS Euler Roman fonts on the system,
                   using the 'upmath' package.^^J}%
       \usepackage{upmath}}
      {\typeout{^^JFound AMS Euler Roman fonts on the system, but you
                   dont seem to have the}%
       \typeout{'upmath' package installed. JPP.cls can take advantage
                 of these fonts, if you use 'upmath' package.^^J}%
      }
  \else
  \fi
\fi

\ifCUPmtlplainloaded \else
  \checkfont{msam10}
  \iffontfound
    \IfFileExists{amssymb.sty}
      {\typeout{^^JFound AMS Symbol fonts on the system, using the
                'amssymb' package.^^J}%
       \usepackage{amssymb}%

      }{}
  \fi
\fi

\ifCUPmtlplainloaded \else
  \IfFileExists{amsbsy.sty}
    {\typeout{^^JFound the 'amsbsy' package on the system, using it.^^J}%
     \usepackage{amsbsy}}
    {\providecommand\boldsymbol[1]{\mbox{\boldmath $##1$}}}
\fi

\newsavebox{\astrutbox}
\sbox{\astrutbox}{\rule[-5pt]{0pt}{20pt}}

\title[Microstructure in perpendicular shocks]{Microstructure in two- and three-dimensional hybrid simulations of perpendicular collisionless shocks}

\author[D. Burgess, P. Hellinger, P. W. Gingell and P. M. Tr\'{a}vn\'{\i}\v{c}ek ]%
{DAVID BURGESS$^1$%
  \thanks{Email address for correspondence: D.Burgess@qmul.ac.uk},\ns
PETR HELLINGER$^{2,3}$\break
PETER W. GINGELL$^1$,
\and PAVEL M. TR\'{A}VN\'{I}\v{C}EK$^{2,3,4}$}

\affiliation{$^1$School of Physics and Astronomy, Queen Mary University of London, London E1 4NS, UK\\[\affilskip]
$^2$Astronomical Institute AS CR, Bocni II/1401, CZ-14131 Prague, Czech Republic\\[\affilskip]
$^3$Institute of Atmospheric Physics, AS CR, Bocni II/1401, CZ-14131 Prague, Czech Republic\\[\affilskip]
$^4$Space Sciences Laboratory, University of Berkeley, 7 Gauss Way, Berkeley, CA 94720, U.S.A.}

\pubyear{XXXX}
\volume{XXXX}
\pagerange{XXX--XXX}
\date{?; revised ?; accepted ?. - To be entered by editorial office}
\begin{document}

\maketitle

\begin{abstract}

Supercritical collisionless perpendicular shocks have an average macrostructure determined primarily by the dynamics of ions specularly reflected at the magnetic ramp. Within the overall macrostructure, instabilities, both linear and nonlinear, generate fluctuations and microstructure. To identify the sources of such microstructure, high-resolution two- and three-dimensional simulations have been carried out using the hybrid method, wherein the ions are treated as particles and the electron response is modelled as a massless fluid. We confirm the results of earlier 2-D simulations showing both field-parallel aligned propagating fluctuations and fluctuations carried by the reflected-gyrating ions. In addition, it is shown that, for 2-D simulations of the shock coplanarity plane, the presence of short-wavelength fluctuations in all magnetic components is associated with the ion Weibel instability driven at the upstream edge of the foot by the reflected-gyrating ions. In 3-D simulations we show for the first time that the dominant microstructure is due to a coupling between field-parallel propagating fluctuations in the ramp and the motion of the reflected ions. This results in a pattern of fluctuations counter-propagating across the surface of the shock at an angle inclined to the magnetic field direction, due to a combination of field-parallel motion at the Alfv\'en speed of the ramp, and motion in the sense of gyration of the reflected ions.
\end{abstract}


\hyphenation{super-critical quasi-perp-end-icular}

\section{Introduction}

For perpendicular collisionless shocks above a certain critical Mach number there is no stable two-fluid solution based solely on resistivity. However, quasi-steady solutions are in fact possible, and such shocks observed, for supercritical shocks because some fraction of the incoming ions are specularly (or near-specularly) reflected at the shock. The gyration of the reflected ions as they return to the shock creates a foot (a gradual decrease in average velocity, and increase in density and magnetic field) upstream of the ramp (where the flow quantities change most rapidly). At the ramp, and for a short distance downstream, the magnetic field increases above the downstream asymptotic value, forming what is termed the overshoot. The gyrating reflected ions provide the downstream temperature increase required by the shock conservation relations. The foot/ramp/overshoot structure of the supercritical perpendicular (and quasi-perpendicular) shock is well established from both observations \citep[e.g.,][]{sckopke:1983,scudal86} and simulations \citep[e.g.,][]{leroal82}; see also \citet{buma15}.

Numerical simulations \citep{biwe72,ques85,leda87,lesa92} revealed a strongly nonstationary behaviour of quasi-perpendicular shocks in the supercritical regime. Similar behaviour is also evidenced with  parameters relevant to the heliospheric and astrophysical contexts \citep{shho00,schmal02}. This nonstationarity is characterized by a periodic self-reformation of the shock front over ion time scales. In the case of strictly perpendicular shocks, the self-reformation process reveals itself to be very sensitive to Alfv\'en Mach number $M_A$ and ion upstream $\beta_\mathrm{i}$ \citep{hellal02,hadaal03}. Recent reviews \citep{hell03,lembal04} stress the importance of the self-reformation process for the shock properties. By using one-dimensional  (1-D) full particle-in-cell (PIC) simulation  code, \cite{schoal03} pointed out that the nonstationary self-reformation process switches off some acceleration mechanisms, for example the shock surfing mechanism, which has been proposed for time-stationary shock solutions \citep{sagd66,liza99}. On the other hand, 1-D PIC simulations \citep{schmal02,leeal04} show that the shock reformation leads to a strong energization of a portion of reflected ions during their subsequent interaction with the nonstationary shock.

The study of nonstationarity is dominated by results based on simulations and theory, which reflects the difficulty of such observation at heliospheric shocks. The difficulties are associated with interpreting differences between multipoint spacecraft measurements when only a few spatial scales can be observed, and the conversion from time to space in the data relies on knowledge of the shock speed relative to the observation points. The separation between macrostructure and microstructure was already considered in an influential two spacecraft study using ISEE data, where it was found that, by using inter-correlation with different averaging periods, the correlation between the two spacecraft decreased strongly when the averaging period was about 0.15 times the upstream proton gyroperiod \citep{scudal86}. The inference is that fluctuations seen at short time scales are associated with nonstationarity, or the breakdown of the usual assumption of one-dimensionality. Using four-point Cluster data, \citet{lobzal07}, presented evidence of shock reformation based on cross-correlation analysis, claiming that the macrostructure as seen by the different Cluster spacecraft could be explained by reformation. Also, the fraction of reflected ions was seen to vary strongly, and bursts of high-frequency waves were seen. The interpretation was that the reformation was driven by a fluctuating reflected fraction, and the generation of non-stationary whistler wave packets was as in the reformation mechanism proposed by \citet{krasnoselskikh:2002}. 

Observations of reformation are usually in the context of some assumed model, which means that interpretations can vary. For example, \citet{comisel:2011} performed 1-D PIC simulations and made comparisons with the same shock crossing studied by \citet{lobzal07}. Based on the simulation results, they concluded that the high frequency fluctuations were due to the modified two-stream instability (MTSI), rather than the whistler gradient catastrophe model of \citet{krasnoselskikh:2002}. It could be noted that the use of a one-dimensional simulation introduces some strong assumptions about the microstructure of the shock, which could affect comparisons with observations.

Two-dimensional hybrid simulations were initially used to study the possible competition between Alfv\'en ion cyclotron (AIC) and mirror instabilities due to the perpendicular ion temperature anisotropy produced by the reflected-gyrating ions \citep{wiqu88}. It was found that the ``ripples'' (surface fluctuations) at the ramp were most likely due to the AIC instability, although no exact agreement with linear homogeneous theory was found. \citet{lobu03}, using similar simulations, found that the shock ramp fluctuations propagated across the shock at the Alfv\'en speed of the shock overshoot, confirming the importance of AIC-like waves. Both these studies used a 2-D simulation plane which was parallel to the shock coplanarity plane, i.e., the upstream magnetic field was in the simulation plane. When the 2-D simulation plane is arranged perpendicular to the shock coplanarity plane (i.e., the upstream magnetic field is perpendicular to the simulation plane) it is found that a new type of shock fluctuation dominates the shock \citep{busc07}. These fluctuations are associated with a modulation of the fraction of reflected ions, and propagate across the shock at a speed, and in a direction, determined by the reflected ions.

The question of non-stationarity at perpendicular and quasi-perpendicular shocks has also been addressed by studies using 2-D simulations. \citet{hellal07} found that shock reformation could be suppressed due to the presence of oblique whistlers, which are not present in 1-D simulations. \citet{lembal09} reported full-particle PIC simulations with reasonably high ion-electron mass ratio of 400, and showed that whistler waves in the foot can also suppress reformation. The 2-D hybrid simulations of \citet{yuanal09}, on the other hand, seemed to show that reformation could occur despite the presence of whistler waves. However, it has been shown that reformation can appear to be suppressed if quantities are averaged over a direction tangential to the shock surface, even though the time evolution at a fixed point on the shock surface does show oscillations attributed to reformation \citep{umeda:2010}.

There are only a limited number of studies using three-dimensional shock simulations \citep[e.g.,][]{thomas:1989,hellinger:1996,giacalone:2000,guo:2013}, so it is still an open question to what extent the behaviour and microstructure seen in 2-D simulations persists in 3-D. There is also the possibility that new types of behaviour emerge in 3-D. Thus in this paper we perform a series of 2-D and 3-D simulations to describe the microstructure see in hybrid simulations of the perpendicular shock. We restrict the study to the hybrid simulation method so as to concentrate on ion-scale behaviour, and to supercritical but moderate Mach number. The latter condition allows the different mechanisms to be studied without excessive cross-coupling. In particular the shocks we study are not dominated by reformation processes, due to the choice of Mach number, ion beta, and the inclusion of a relatively strong resistivity. The hybrid simulation technique is also increasingly being used to model astrophysical shocks and particle acceleration, and therefore it is important to understand the physical mechanisms responsible for shock fluctuations and microstructure in such simulations. In this paper we demonstrate the role of the ion Weibel instability as a driver of whistler-like fluctuations in the foot of the shock. The ion Weibel instability has received attention as crucial for the formation of shocks in unmagnetized (or low magnetization) plasmas \citep{spitkovsky:2008}, but here we show that it also plays a role in magnetized shocks as found at the Earth's bow shock and in other heliospheric environments. Furthermore, we show that the three-dimensional microstructure is dominated, at ion scales, by a combination of the shock rippling and reflection fraction instability seen in two-dimensional simulations made in, and perpendicular to the coplanarity plane. The result is a pattern of fluctuations propagating across the shock surface at an angle to both the magnetic field direction and the direction of motion of the reflected-gyrating ions.

We use two- and three-dimensional hybrid simulation codes \citep{matt94}, where electrons are considered as a massless, charge neutralizing adiabatic fluid, whereas ions are treated as macro-particles. The 2-D simulation box is $600\times 512$ with a spatial resolution $\Delta x=0.1c/\omega_{pi}$ and $\Delta y=0.2c/\omega_{pi}$, and there are $1024$ particles per cell in the upstream region. The 3-D simulation box is $512\times128\times 128$ points with a spatial resolution $\Delta x=0.1c/\omega_{pi}$ and $\Delta y=\Delta z=0.2c/\omega_{pi}$, and there are $256$ particles per cell in the upstream region.

The time step for the particle advance is $\Delta t=0.005/\Omega_\mathrm{i}$ whereas the magnetic field is advanced with $\Delta t_B = \Delta t/10$. In these definitions, $c$ is the speed of light, $\omega_{pi}$ is the upstream ion plasma frequency, $\lambda_\mathrm{i}$ is the ion inertial length, and $\Omega_\mathrm{i}$ is the upstream ion cyclotron frequency. Protons and electrons have initial (upstream) ratios between the particle and magnetic pressures $\beta_\mathrm{i}=0.2$ and $\beta_\mathrm{e}=0.01$, respectively. In the generalized Ohm's law we use a small resistivity $\eta = 10^{-2} \mu_0 v_\mathrm{A}^2/\Omega_\mathrm{i}$; here $\mu_0$ is the magnetic permeability of vacuum and $v_A$ is the upstream Alfv\'en velocity. For 2-D and 3-D simulations the plasma is streaming along the $x$ axis (protons are continously injected at $x=0$) and interacts with a piston at $x_\mathrm{max}$ which launches a shock travelling towards negative $x$.

\section{Results: 2-D and 3-D hybrid simulations}

\subsection{2-D hybrid simulation in the coplanarity plane\label{sect:2dinplane}}

\begin{figure}
\centerline{\includegraphics[width=13cm]{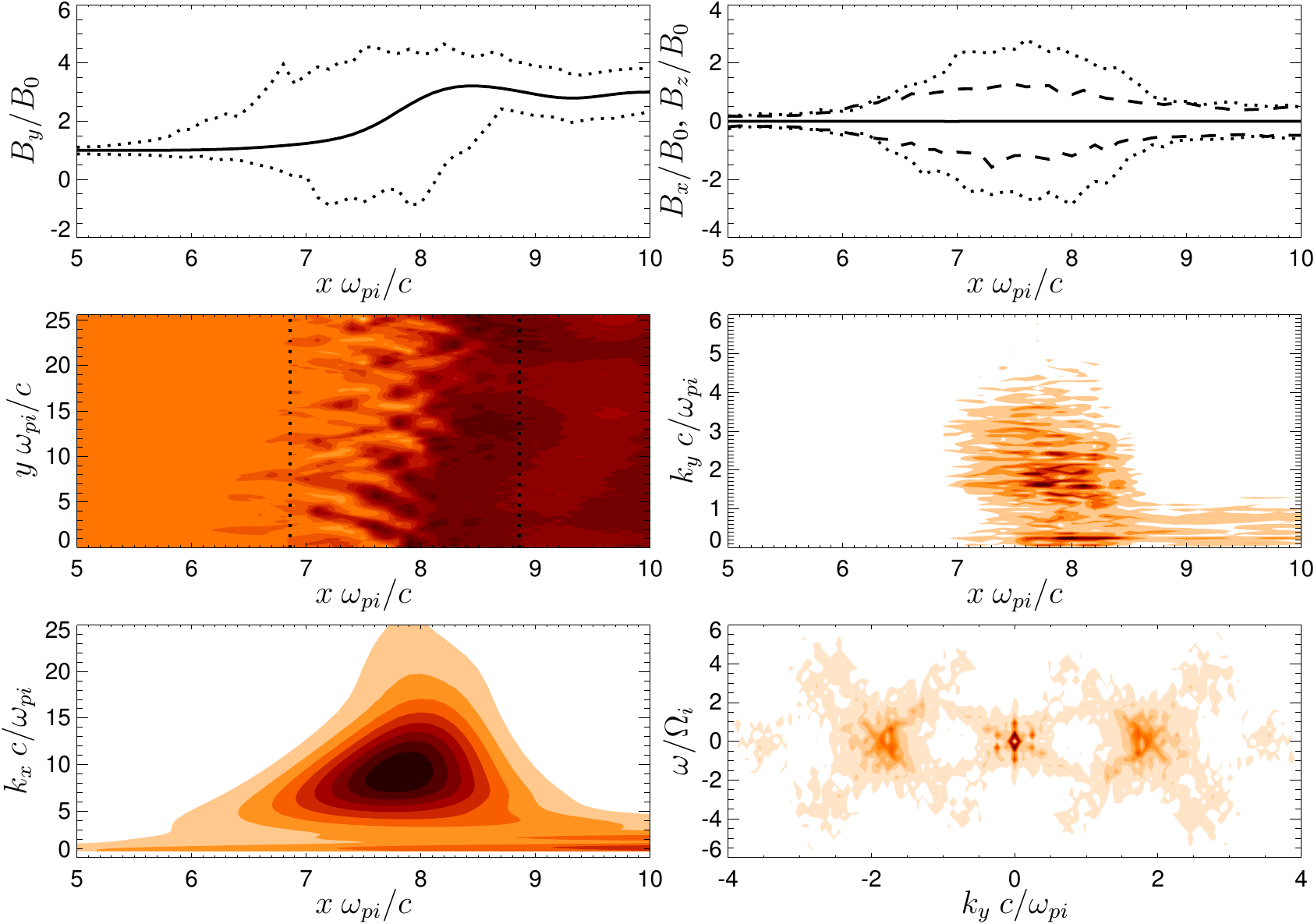}}%
\caption{2-D hybrid simulation in the coplanarity plane:
(top left) the average (see the text) profile of $B_y$ (solid line) 
and the two extreme values of $B_y$ (dotted lines);
(top right) average profiles  of $B_x$ and $B_z$ (solid line),
the dotted and dashed lines show the two extreme values of $B_z$
and $B_x$, respectively.
(middle left) The colour scale plot of $B_y$ at $t=40\Omega_\mathrm{i}^{-1}$ as a function of $x$ and
$y$; (middle right) the  colour scale plot of the fluctuating magnetic field
$\delta B$ at $t=40\Omega_\mathrm{i}^{-1}$ as a function  of $x$ and $k_y$. 
(bottom left) the  colour scale plot of the fluctuating magnetic field
$\delta B$ at $t=40\Omega_\mathrm{i}^{-1}$ as a function  of $x$ and $k_x$
(using the wavelet transform). (bottom right) The fluctuating magnetic field
$\delta B$ between $20$ and $40$ $\Omega_\mathrm{i}^{-1}$ (within the shock front)
as a function  of $k_y$ and $\omega$ in the average shock rest frame; here the
mean shock profile has been removed.
}
\label{fig:f1}
\end{figure}

We first consider a 2-D simulation of a perpendicular shock with $\theta_{Bn}=90^\circ$, and $M_A=3.3$, such that the upstream magnetic field is in the simulation plane, meaning that the simulation plane is also the coplanarity plane. Animations of the magnetic field fluctuations in the foot show a complex behaviour of the phase fronts. At the ramp and overshoot field-parallel propagating waves/structures are seen corresponding to the AIC modes previously studied \citep{wiqu88,hellinger:1997,lobu03}. Also present are shorter wavelength fluctuations with obliquely oriented phase fronts (neither field-aligned nor aligned with the shock normal) which seem to have properties varying between propagating and phase standing. In figure \ref{fig:f1} we show an overview of the shock fluctuations with a spectral analysis. The envelope of magnetic fluctuations is shown in the top panels, demonstrating, as earlier works have shown, that the amplitudes of fluctuations in all components of the field are comparable. This makes impossible any analysis based on simplistic expectations of mode polarization. (For example, the AIC is transverse and so would be expected to be seen only in the $B_x$ and $B_z$ components.)

A snapshot of the field magnitude fluctuations shows (figure \ref{fig:f1}, middle left) the complex pattern of phase fronts at one time ($t_m=40 \Omega_\mathrm{i}^{-1}$). Only a small section of the simulation is shown, whereas the $k$-space and $\omega-k$ analysis uses the full domain. The average profile in $x$ has been removed. Spectral analysis (wavevector) analysis of this pattern using a wavelet transform in $x$ and Fourier transform in $y$ is shown in the middle right and bottom left panels of figure \ref{fig:f1}. As expected the field magnetic fluctuation power across wavevector spaces $k_x$ and $k_y$ peaks around the shock ramp at $x\sim8$. The dependence of power with $k_x$, especially the spread of power at $k_x \sim 5$ is dominated by the wavelet response to the finite size of the foot. On the other hand, the dependence on $k_y$ show peaks at small $k_y$ (apparently close to zero), and at $k_y\sim 2$.

The latter structure is analyzed in the $\omega-k_y$ power spectrum shown in the bottom right panel of figure \ref{fig:f1}. This is the result of a two-dimensional time-space transform of the variation of field magnitude within the shock front as a function of $y$ between $t=20$ and $40 \Omega_\mathrm{i}^{-1}$, and in the shock rest frame with the mean shock profile removed. Recall that the power spectrum has two-fold symmetry with $P(-\omega,-k_y) = P(\omega,k_y)$. The spectrum consists of three components: The first is centred at $k_y=0$, with enhancements across a range of frequencies, and corresponds to fluctuations in the mean shock speed, and mean shock profile. Such variations are a common feature of hybrid simulations, as noted in early 1-D studies \citep{leroal82}. The second component has the signature of propagating waves with phase speed $\omega/k_y \approx \pm 2 V_\mathrm{A}$. This component corresponds to the field-parallel fluctuations propagating at approximately the Alfv\'en speed of the shock ramp/overshoot as reported earlier \citep{lobu03}. The third component in the power spectrum is an enhancement centred on $\omega=0$ and $k_y \approx \pm 2 $, but also with the signature of propagating fluctuations with a slope $\omega/k_y \approx \pm 4 V_\mathrm{A}$. Visually the component can be described as a cross in $\omega-k_y$ shifted to non-zero $k_y$. This second component can be explained as the sidebands produced by the modulation of a propagating signal by a zero-frequency, $k_y\sim 2$ signal. Although the propagating signal might be tentatively identified with oblique whistler mode fluctuations (seen in the middle left panel of figure \ref{fig:f1}), a source for a zero frequency, $k_y\approx 2$ signal is not obvious.

We next identify the zero-frequency $k_y\approx 2$ component in the field fluctuations, and then explore the hypothesis that it is due to the ion Weibel instability. As discussed earlier, there are several coupled systems at work within the foot which can make it difficult to disentangle cause and effect. For example, an important factor is that at any point in the ramp the reflected fraction depends on the fields at that point. But the reflected fraction modulates the foot (upstream of the ramp) with a delay determined by the particle trajectories. Thus, considering spatial variation across the shock surface (i.e., in the $y$ direction), fluctuations at the ramp, e.g., due to AIC-like fluctuations, can spatially modulate the density of reflected ions at the upstream edge of the foot. In order to eliminate this particular source of modulation of the foot, we carry out simulations which create, for a short time, the idealized case of uniform reflected ion density across the shock front. This allows us to identify fluctuations driven by instabilities of the reflected ion beam, without the complication of spatial variation of beam density. The method is to carry out a two-dimensional simulation, but to split the simulation domain into two parts. In the first part (nearest the wall which launches the shock) the shock is constrained to behave as if it were one dimensional by, for each field componennt and at each $x$ position, averaging over the $y$ direction and then forcing the fields to take that average value for all $y$. Any variation with $y$ (transverse to the shock normal) is suppressed, and the shock structure depends only on $x$. However, in the second (left-hand) part of the simulation domain, the full two-dimensional system is allowed to evolve as normal. The shock is launched at the right-hand wall and travels to the left. Thus, the shock is initially one-dimensional with no variation across the $y$ direction. However, the shock will gradually move into the left-hand part where two-dimensional structure can develop. It will be firstly the upstream edge of the foot, then the foot itself, followed by the ramp which progressively is is allowed to develop two-dimensional structure. Of course, there is some loss of self-consistency at the boundary between 1-D and 2-D subdomains, but by following the evolution until the shock is fully 2-D it is apparent that any artefacts which have been introduced are limited to only a few cells.

\begin{figure}
\centerline{\includegraphics[width=10cm]{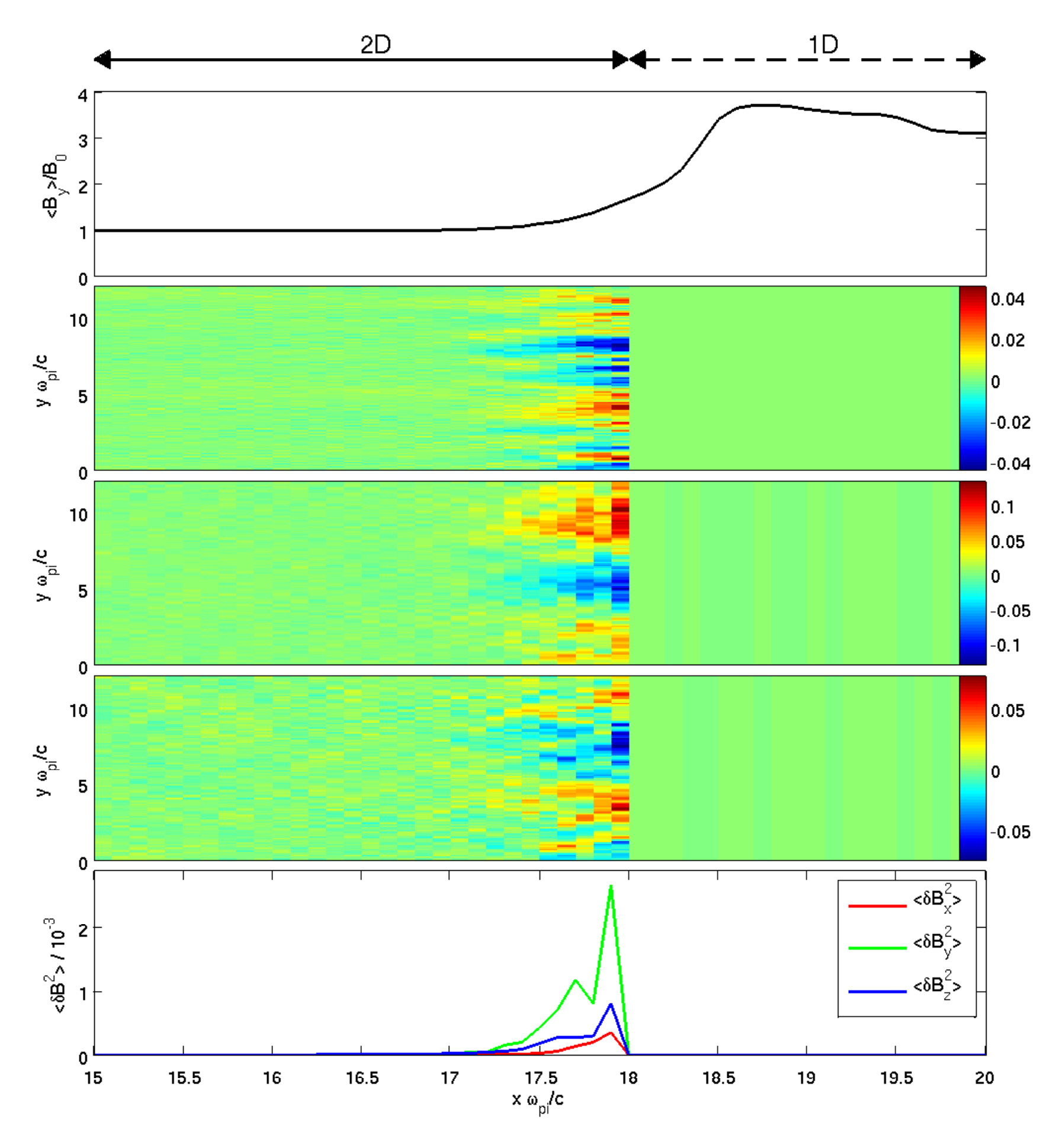}}%
\caption{2-D hybrid simulation in the coplanarity plane with a forced 1-D to 2-D transition: (top panel) $B_y$ profile averaged over $y$ direction, with 1-D and 2-D regions indicated; (middle panels) colour scale plots of fluctuations in $B_x$, $B_y$ and $B_z$ (from top to bottom) after removal of $y$-averaged profile; (bottom panel) power in component fluctuations as a function of $x$. Only a small region of the simulation domain, around the shock transition, is shown. 
}
\label{fig:f2}
\end{figure}

This technique of transitioning the shock foot from one- to two-dimensional is illustrated in figure \ref{fig:f2} which shows the $y$-averaged field magnitude profile, the magnetic field component fluctuations as a function of $x$ and $y$, and the $y$-averaged profile of the power in the components. The $y$ range in this simulation is twice that shown, and only a small region around the shock in the $x$ direction is shown. The regions which are treated as 1-D and 2-D are marked. It can be seen that the shock has propagated far enough that the front edge of the foot is allowed to behave two-dimensionally, whereas the ramp at $x=18.3$ is still constrained to be one-dimensional. The density of reflected ions at this time has been confirmed to be uniform in $y$ across the foot. Fluctuations in all field components grow extremely rapidly almost immediately as the foot moves into the 2-D subdomain. The $B_x$ component shows a wave pattern with wavevector in the $y$ direction and wavelength of approximately $3 \lambda_\mathrm{i}$ (i.e., $k\lambda_\mathrm{i}\approx2$). The $B_y$ and $B_z$ components show slightly different behaviour, with a more complicated pattern which seems to develop further downstream of the front of the foot. The power in the $B_y$ and $B_z$ components is higher than in $B_x$, but all components develop in a similar way across the front edge of the foot.

\begin{figure}
\centerline{\includegraphics[width=12cm]{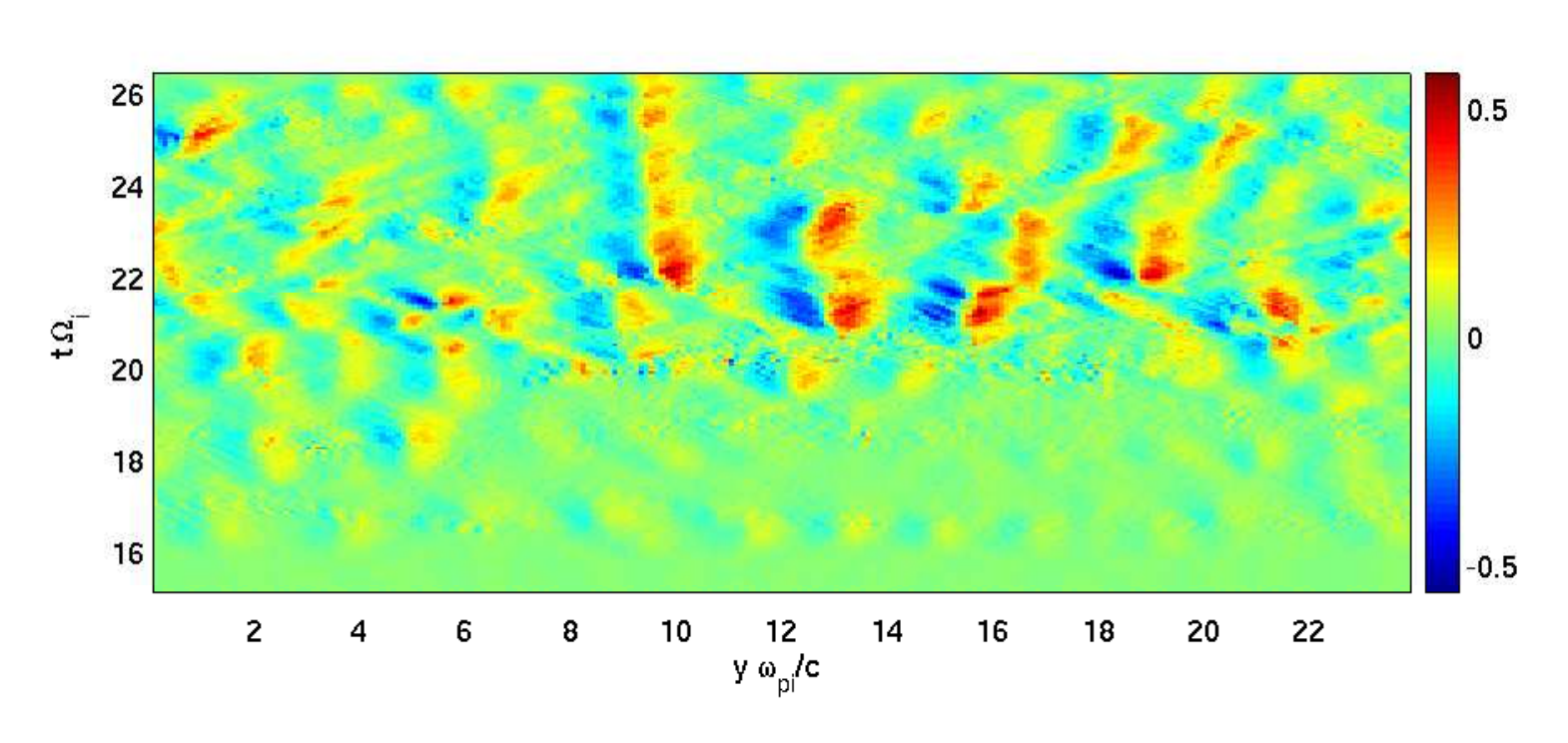}}%
\caption{Evolution of $B_x$ as a function of $y$ and time, for the shock experiment shown in figure \ref{fig:f2}, at a fixed distance upstream of the shock ramp. At early times the shock at this position transitions from 1-D to 2-D, and at later times the entire foot and ramp have fully-developed 2-D structure.
}
\label{fig:f3}
\end{figure}

The $B_x$ component of the fluctuations at the front (upstream) edge of the foot seems to have the appropriate value of $k_y\approx 2$ for the modulation seen in the $\omega-k_y$ power distribution in figure \ref{fig:f1}. But that modulation also requires $\omega \approx 0$. To investigate the time behaviour in figure \ref{fig:f3} we plot the time evolution of the $B_x$ fluctuations as a function of time and $y$, at a fixed distance upstream of the ramp in a frame in which the shock is at rest. The time range covers that corresponding from when the foot first enters the 2-D simulation subdomain until when all the foot and ramp are fully 2-D. At the earliest time (after $\Omega_\mathrm{i} t= 16 $) a very clean monochromatic wave is seen with wavelength of about $3 \lambda_\mathrm{i}$. There appears to be little propagation, so the frequency is zero, or close to zero. At slightly later times, after $\Omega_\mathrm{i} t= 19 $, the time evolution becomes more complex and more variable across $y$. It appears that the pattern is still dominated by fluctuations with $\omega\sim 0$. The density of reflected ions at this position is uniform across $y$ at early times, but strongly variable once the pattern in $B_x$ becomes more variable. It can be noted that the timescale for a reflected ion to return to the shock is approximately $1.9\Omega_\mathrm{i}^{-1}$, and the flow transit time is roughly $\Omega_\mathrm{i}^{-1}$ so the time period taken for the reflected ion density to become nonuniform over $y$ is consistent with modulation of the reflected fraction by the fluctuations seen at earlier times, once they convect into the ramp.

\begin{figure}
\centerline{\includegraphics[width=12cm]{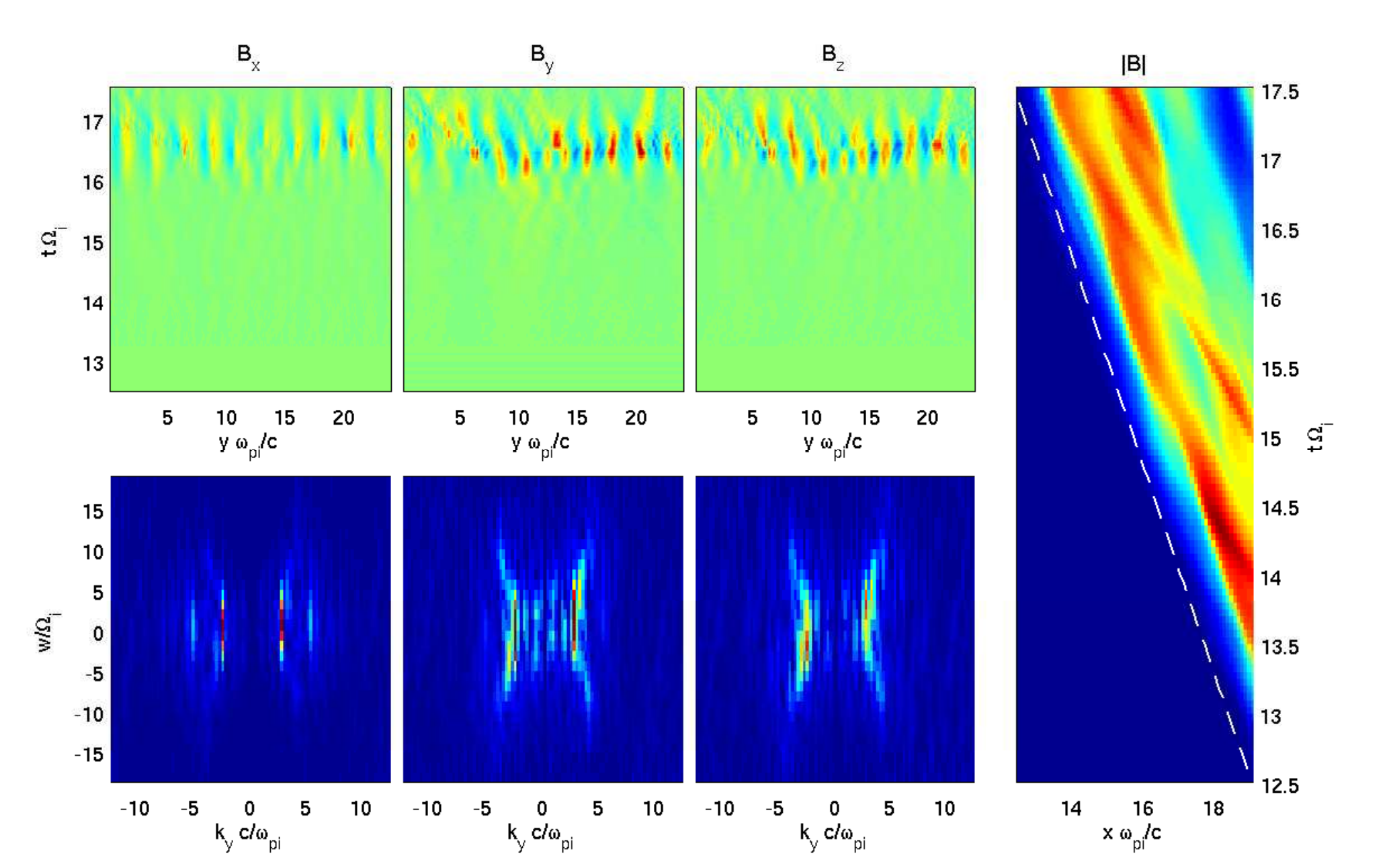}}%
\caption{Evolution of $B_x$, $B_y$ and $B_z$ (top panels) as a function of $y$ and time, for the 2-D hybrid simulation shown in figure \ref{fig:f2}, at a fixed distance upstream of the shock ramp (as shown in the right-hand panel). Frequency - wave vector power distributions for $B_x$, $B_y$ and $B_z$ (bottom panels) for the fluctuations shown in the top panels.
}
\label{fig:f4}
\end{figure}

We can further distinguish the behaviour of the fluctuation components in figure \ref{fig:f4} which shows the space-time and also $\omega-k_y$ plots taken at position at a fixed distance upstream of the shock ramp in the shock rest frame, as shown in the right-hand panel. The sampling position is very near the upstream edge of the shock, and the time period covers the emergence of the shock front part of the foot into the 2-D subdomain. The $B_x$ fluctuations show a clear non-propagating signal, confirmed by the $\omega-k_y$ behaviour, with a strong signal at $k\lambda_\mathrm{i}\approx2$. The $B_y$ and $B_z$ components are, again, more complex, but the $\omega-k_y$ plot shows a mixture of zero frequency and propagating characteristics shifted in $k_y$.

Our interpretation of these results is as follows: At the very upstream edge of the foot, the reflected ions, near their turnaround point, can drive an instability with a zero-frequency, relatively high $k_y$ wave pattern, with wave vector in the field-parallel $y$ direction. The signal produced is most clear in $B_x$, but is coupled to $B_y$ and $B_z$ which produces propagating (nonzero frequency) waves. The propagating waves are essentially the fluctuations which would be identified as ``whistler fluctuations'' since they are driven at high $k_y$ and have corresponding relatively high phase speeds. The coupling between the field components is not straightforward, and it is complicated by the strong inhomogeneity in field and velocity as the plasma decelerates in the foot and ramp. In a shock with a fully 2-D developed structure the fraction of reflected ions is non-uniform over $y$ due to fluctuations at the ramp, which in turn may be due to AIC fluctuations generated there, or fluctuations coupled to the foot. Despite the non-uniform density of the reflected ion beam at the upstream edge of the shock, the zero-frequency, high $k_y$ instability can still operate. The wave pattern from this instability drives (propagating) whistler fluctuations, and it is this modulation by the zero-frequency component that is responsible for the power contribution shifted to $k\lambda_\mathrm{i}\approx\pm2$, as seen in the field magnitude at the ramp (figure \ref{fig:f1}, bottom right-hand panel).

An analogy is the pattern of surface fluctuations produced by a linear grid of fine wires perpendicular to a steady fluid flow, where the wires are fine enough to generate dispersive ripples. Very close to the grid (relative to the inter-wire spacing) the $\omega-k$ spectrum transverse to the flow will mainly show the spatial frequency of the grid spacing. Further downstream the spectrum will show propagating modes associated with the ripples, but modulated by the spatial frequency of the grid. Far downstream, interaction between waves from many grid wires may produce a more complex, even turbulent, state.

In the 1-D/2-D simulations the rapid appearance of the fluctuations at the upstream edge as it enters the 2-D subdomain indicates that they are driven by an instability of the reflected ion beam. Consider specularly reflected ions at the perpendicular shock. At that point, where the reflected ions turn around, their $x$ velocity is zero, and they form a beam in velocity space perpendicular to the magnetic field with $v_z \approx \sqrt{3} M_\mathrm{A} V_\mathrm{A}$. For this situation it is known that an instability due to a cross-field ion drift \citep{chang:1990,wu:1992} can drive an electromagnetic, zero-frequency wave with a high growth rate. The mechanism for this instability is essentially that of the Weibel instability \citep{weibel:1959} as described by \citet{fried:1959}. The instability mechanism is usually considered for electrons, but it is also viable for ions, and so in this context we follow \citet{chang:1990} who used the term ``ion Weibel instability.''

For the homogeneous, cold ion beam case, with assumptions similar to the hybrid simulation (see Appendix), there is instability at zero frequency for all $k$, with a maximum growth rate for $\boldsymbol{k}$ aligned field-parallel. The growth rate $\gamma$ decreases at small $k$, and has a typical value of approximately $\gamma \sim 2.2 \Omega_\mathrm{i} $, for beam density $n_\mathrm{b}=0.2 n_\mathrm{i}$ and cross-field velocity $v_\mathrm{b}=5 V_\mathrm{A}$. An estimate of a typical value of $k$ for which the growth rate is close to its maximum is approximately $k_\mathrm{m} \sim  2.2 $ for the simulation shown here, which is close to the value seen in the results. The simplifications used in these estimates are discussed in the Appendix.

\subsection{2-D hybrid simulation perpendicular to the coplanarity plane}

We have presented evidence that, in 2-D hybrid simulations, some of the microstructure which is ``whistler-like'' is due to driving from nonpropagating waves associated with the ion Weibel instability at the upstream edge of the ion foot. It is an important question whether this scenario is valid in the case where the shock is allowed to have three-dimensional spatial structure. In this section, as an intermediate case, we consider a 2-D hybrid simulation, similar to that shown in figure \ref{fig:f1}, but with the simulation plane perpendicular to the shock coplanarity plane. This corresponds to the upstream magnetic field being arranged perpendicular to the simulation plane. This configuration was studied by \citet{busc07} who found a nonlinear instability associated with the reflected ion fraction. The pattern formed in this instability is carried by the specularly reflected ions, and thus travels with their speed in the cross-field direction.

\begin{figure}
\centerline{\includegraphics[width=13cm]{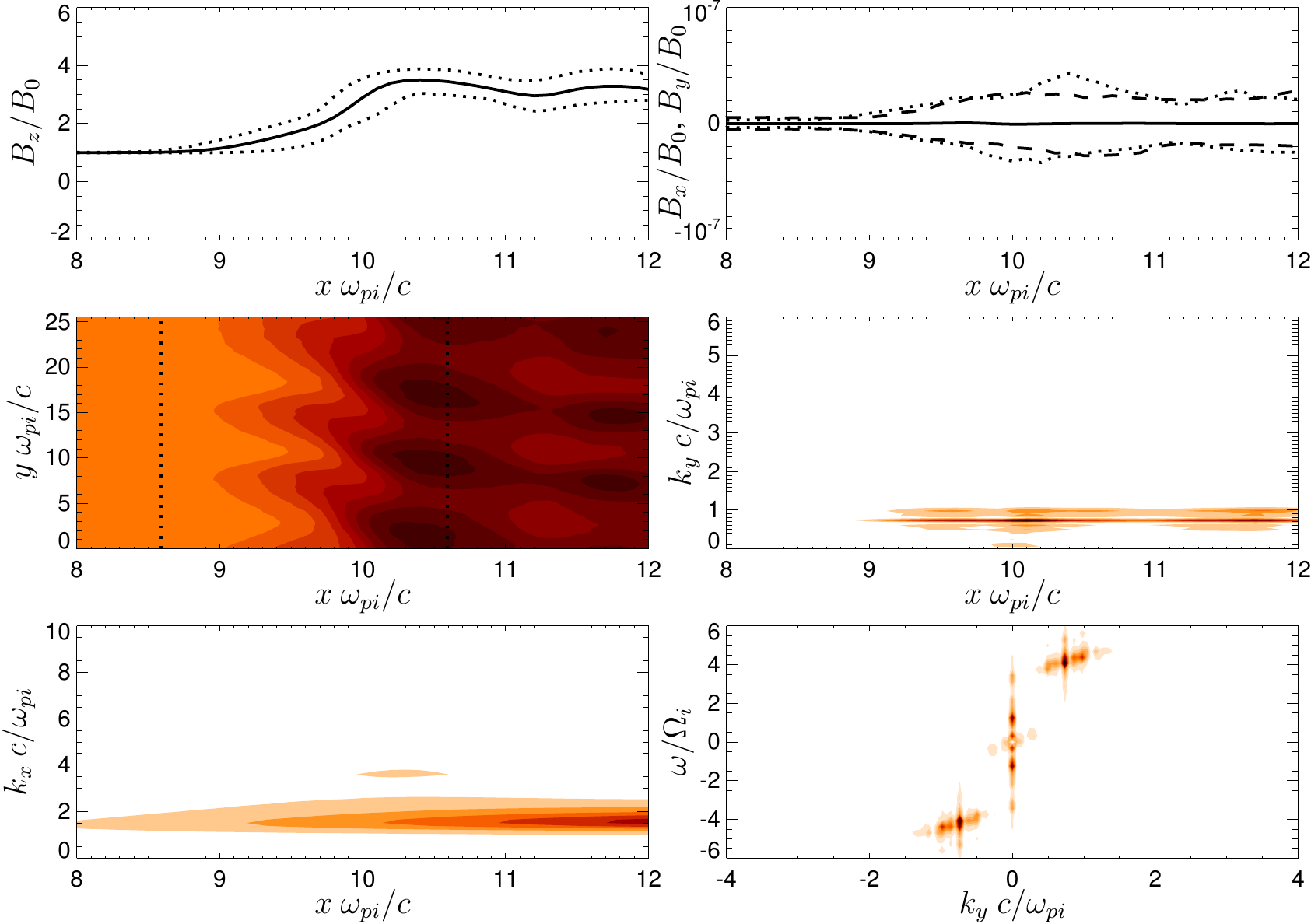}}%
\caption{2-D hybrid simulation with simulation plane perpendicular to the shock coplanarity plane: (top left) the average (see the text) profile of $B_z$ (solid line) and the two extreme values of $B_z$ (dotted lines); (top right) average profiles  of $B_x$ and $B_y$ (solid line), the dotted and dashed lines show the two extreme values of $B_y$ and $B_x$, respectively. (middle left) The colour scale plot of $B_z$ at $t=40\Omega_\mathrm{i}^{-1}$ as a function of $x$ and $y$; (middle right) the  colour-scale plot of the fluctuating magnetic field $\delta B$ at $t=40\Omega_\mathrm{i}^{-1}$ as a function  of $x$ and $k_y$. (bottom left) the  colour-scale plot of the fluctuating magnetic field $\delta B$ at $t=40\Omega_\mathrm{i}^{-1}$ as a function  of $x$ and $k_x$ (using the wavelet transform). (bottom right) The fluctuating magnetic field $\delta B$ between $20$ and $40$ $\Omega_\mathrm{i}^{-1}$ (within the shock front) as a function  of $k_y$ and $\omega$ in the average shock rest frame; here the mean shock profile has been removed.
}
\label{fig:f5}
\end{figure}

In figure \ref{fig:f5} we show show the results of a simulation with parameters similar to the shock shown in figure \ref{fig:f1}, where the upstream magnetic field is in the $z$ direction, and the simulation plane is the $x-y$ plane. Comparing with figure \ref{fig:f1} it is evident that the behaviour of fluctuations in the foot/ramp is very different compared to when fluctuations with a field-parallel orientation are allowed. The overall level of fluctuations is much lower (top panels), the $k_y$ and $k_x$ spectra of fluctuations are very different from the earlier case, with negligible power at high $k$. In the $k_y$ spectrum there is only a single, strong component at $k_y\lambda_\mathrm{i}\approx 0.75$. From the snapshot of field magnitude (middle left-hand panel) this is seen to take the same form as the nonlinear instability found by \citet{busc07}, i.e., a pattern of ``tongues'' of field magnitude (and also reflected ion density) in the foot. The $\omega-k_y$ power spectrum (bottom right-hand panel) shows a component spread over $\omega$ but at zero $k_y$. This probably corresponds to small variations in the shock speed, independent of $y$, which are not taken into account when translating into the shock frame assuming a constant shock speed. The only other component corresponds to peak power at $k_y\approx0.75$ and $\omega\approx 4.1$, or a phase speed of 5.5 $V_\mathrm{A}$. The cross-field velocity of the reflected ions at their turnaround point is $1.7 M_\mathrm{A} V_\mathrm{A}\approx 5.6$, so the phase velocity of the pattern seen in the field magnitude, matches the cross-field velocity of the reflected ions at their turnaround point, in agreement with \citet{busc07}.

In the context of the current study, excluding fluctuations with a field-parallel orientation (by the choice of simulation plane) suppresses all of the mechanisms for microstructure discussed in section \ref{sect:2dinplane}. Instead, a strong and monochromatic variation of the foot/ramp is found associated with the cross-field motion of the reflected ions.

\subsection{3-D hybrid simulation}

In order to determine which of the types of microstructure described above survive or dominate when full three-dimensional spatial dependence is allowed, we have performed a 3-D hybrid simulation of a shock with a similar Mach number ($M_A=3.1$) and otherwise identical parameters as for the earlier simulations. The upstream magnetic field is aligned with the $y$ axis, $\boldsymbol{B}=(0,B_0,0)$, so that the coplanarity plane is the $x$-$y$ plane. Specularly reflected ions, from their sense of gyration, will have a $-v_z$ velocity at their turnaround point.

\begin{figure}
\centerline{\includegraphics[width=13cm]{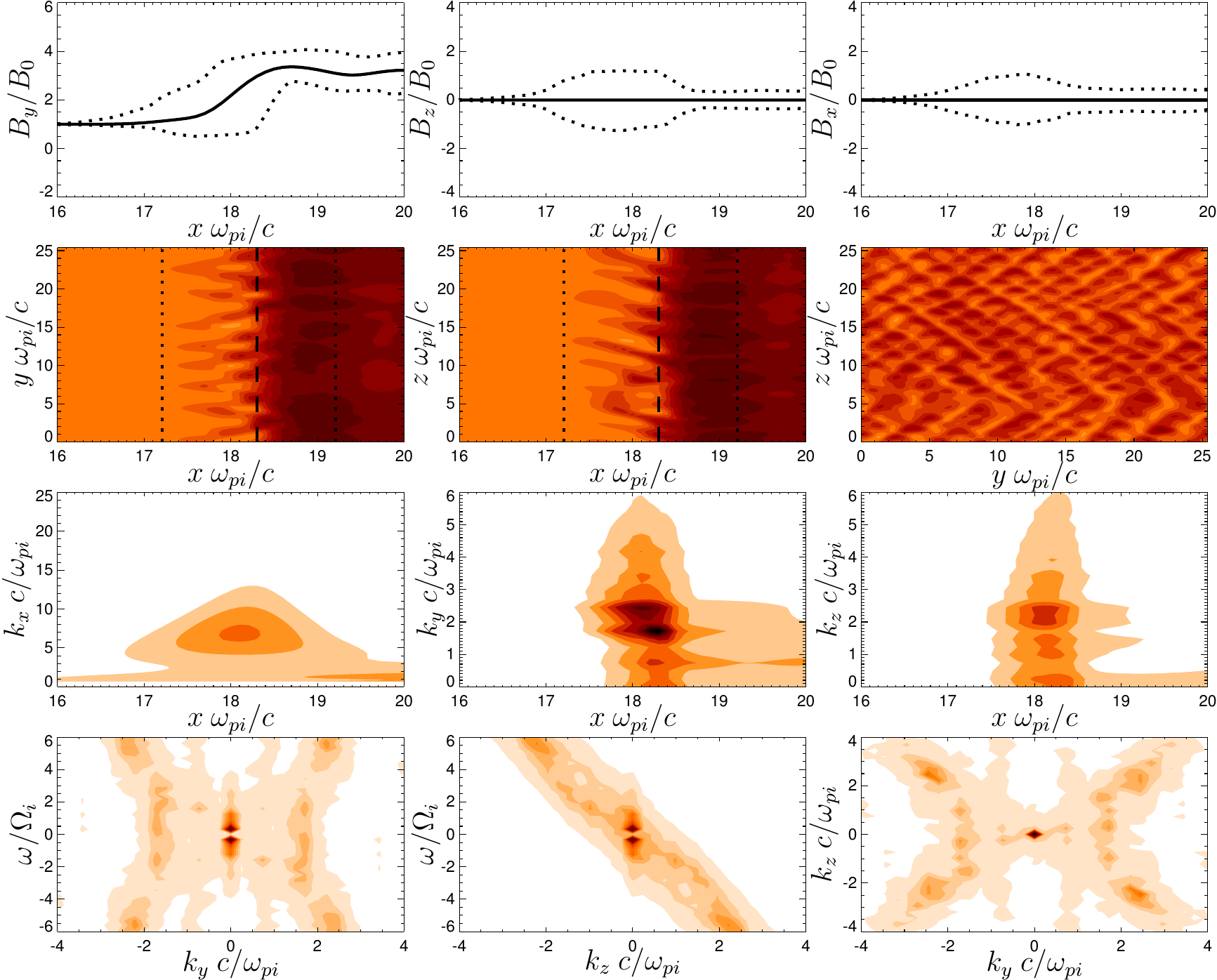}}%
\caption{3-D hybrid simulation $M_A=3.1$:
Top panels show (solid) the average profiles and (dotted) the extreme values of
(left) $B_y$, (middle) $B_z$, and (right) $B_x$ as functions of $x$.
Top middle panels show colour scale 2-D cuts of $B_y$  $t=30\Omega_\mathrm{i}^{-1}$
(left) in $x$-$y$ plane, (middle) in $x$-$z$ plane, and 
(right) in $y$-$z$ plane.
Bottom middle panels show $\delta B$ at $t=30\Omega_\mathrm{i}^{-1}$ as a function
 of (left) $x$ and $k_x$, (middle) $x$ and $k_y$, and (right) $x$ and $k_z$.
Bottom panels show the spectra of $\delta B$ between $10$ and $30$ $\Omega_\mathrm{i}^{-1}$ (within the shock front)
calculated in the average shock rest frame with mean shock profile removed:
(left) $\delta B$ as a function  of $k_y$ and $\omega$,
(middle) $\delta B$ as a function  of $k_z$ and $\omega$,
and (right) $\delta B$ as a function  of $k_y$ and $k_z$.
}
\label{fig:f6}
\end{figure}

The shock microstructure is shown in figure \ref{fig:f6}. In order to facilitate comparisons with the 2-D simulations, we refer to the 2-D simulations as  ``simulation plane parallel to shock coplanarity plane'' (SP$\|$CP, figure \ref{fig:f1}), and ``simulation plane perpendicular to shock coplanarity plane (SP$\perp$CP, figure \ref{fig:f5}).

Firstly, the envelopes of the field components fluctuations (top panels, figure \ref{fig:f6}) are most similar to the  SP$\|$CP case, but the $B_z$ component (i.e., perpendicular to coplanarity plane) has a much reduced level, with $\Delta B_z /B_0 \sim 1$ compared to $\Delta B_z /B_0 \sim 2$. The top middle panels show colour-scale 2-D cuts of $B_y$ at $t=30\Omega_\mathrm{i}^{-1}$ in the $x$-$y$ plane (left), in $x$-$z$ plane (middle), and in $y$-$z$ plane (right). The $x$-$y$ and $x$-$z$ planes can be compared to the middle left panels of figures \ref{fig:f1} and \ref{fig:f5}, respectively, since these are parallel and perpendicular to the coplanarity plane. There are similarities in both cases. For the $x$-$y$ plane (coplanarity plane) there is a pattern of oblique phase fronts, which seems to originate from the leading edge of the foot, where the wavelength is $\sim 3 \lambda_\mathrm{i}$. In the case of the $x$-$z$ plane (perpendicular to coplanarity plane) there is (as for the 2-D SP$\perp$CP simulation) a pattern of ``tongues'' of field increase, but the dominant wavelength is shorter than in the 2-D simulation case. The pattern in the $y$-$z$ plane (essentially across the face of the shock) is remarkably regular, and the maxima form a centred square lattice with some dislocations.

The $k$-space representation for the magnetic field magnitude fluctuations are shown in the bottom middle panels. The $k_x$ profile is similar to the SP $\|$ CP simulation, but with reduced overall magnitude. The $k_y$ profile shows enhancements in the foot between $k_y \sim 1.5 - 2.7$ and a weaker component at higher values only in the foot. The $k_z$ profile against $x$ shows a somewhat similar pattern, but at lower amplitude.

The $\omega-k$-space representations are shown in the bottom panels, taken at $x=18.3$ near the ramp. The $\omega-k_z$ plot shows that the pattern is dominated by a $z$ phase speed of $\approx -2.5 V_\mathrm{A}$. This phase speed is in the same sense as the gyration velocity of the reflected ions at turnaround, but somewhat less than what would be expected from the velocity at the ion turnaround position due to specular reflection. The $\omega-k_y$ plot shows components again with a range of $k_y$ but the phase speed is now $\approx \pm2.5 V_\mathrm{A}$. There is some evidence of a component with $k_y\lambda_\mathrm{i}\approx 1.75$ across a range of $\omega$. This is consistent as a signature of the high-$k$ (zero frequency) modulation pattern seen in the SP $\|$ CP simulation, although there is no indication of whistler-like fluctuations modulated by this signal. The $k_y$-$k_z$ plot (bottom right panel) shows a cruciform pattern as expected from the $\omega-k$-space together with the $k_y\sim 1.75$ (constant) component.

Combining the information from these plots, we infer that the real-space pattern in $y$-$z$ (top middle right panel) comprises, across a range of $k$, two patterns propagating at right angles to each other with $\boldsymbol{k}\approx 2.5(\pm\widehat{\boldsymbol{y}} -  \widehat{\boldsymbol{z}})$. Since the $k_y$ (field-parallel) power indicates propagating modes in both directions, the obvious inference is that it is associated with the AIC-like ripples seen in the SP $\|$ CP simulation, and the phase speed is similar, i.e., the Alfv\'en speed of the shock ramp or overshoot. Similarly, the $k_z$ power shows a phase speed in a single direction, in the sense of gyration of reflected ions, and so it is natural to associate this with the pattern produced by the reflected ions in the SP$\perp$CP simulations. However, since the phase speed is not exactly the same as in that case, being approximately $-2.5 V_\mathrm{A}$ rather than the approximately $-5.3 V_\mathrm{A}$ expected from specular reflection. Inspection of the ion velocity space distributions in the foot of the shock indicate that the lower $v_z$ value of $-2.5 V_\mathrm{A}$ corresponds to the gyrating ions on the part of their trajectory soon after reflection, rather than at the turnaround point. This is consistent with the pattern of fluctuations (i.e., in the $y$-$z$ plane) being a combination of field-parallel propagation at the ramp/overshoot and motion of the reflected ions at the same location. The inference is that the currents responsible for what has been called the AIC-like fluctuations are carried by the reflected ions, and so the pattern of fluctuations propagates in the sense of the reflected ions.

\section{Conclusions}

We have carried out a series of high-resolution hybrid simulations to better understand the sources of microstructure and fluctuations in the foot and ramp of perpendicular shocks. We are here concerned with supercritical shocks where specularly reflected ions dominate the foot, ramp and overshoot structure, and provide the bulk of the ion heating. We have chosen parameters such that the shock Mach number is relatively small but still supercritical, so that nonlinear processes such as reformation are not important. Similarly the chosen parameters (perpendicular shock geometry and value of resistivity) do not favour the presence of dispersive whistlers generated at the shock ramp and then propagating upstream into the foot. The 3-D simulations we have carried out are at higher resolution in space and time than previous works, and also use a larger simulation domain. This has allowed us to use spectral analysis to characterize the microstructure fluctuations.

Using both 3-D and 2-D simulations in different configurations has enable us to separate the contributions from different mechanisms. The hybrid method simplifies the electron response, so that short length scale, high frequency microstructure is not accurately modelled. On the other hand, full particle simulations are currently not possible with realistic plasma parameters over the length and time scales achievable in hybrid simulations. This leads to hybrid simulations being more widely used to study other phenomena such as particle acceleration or global systems, and therefore it is important to have a firm understanding of the mechanisms creating microstructure and fluctuations in hybrid shock simulations.

Two-dimensional simulations of a perpendicular shock with the simulation plane parallel to the shock coplanarity plane show field-parallel structure consisting of: (a) ripples at the shock ramp which propagate across the shock surface parallel to the magnetic field at approximately the Alfv\'en speed of the overshoot \citep{wiqu88,lobu03}; and (b) whistler-like fluctuations with short wavelengths \citep{hellal07,lembal09}. In addition, we have shown that the whistler-like fluctuations are modulated by a non-propagating mode at the upstream edge of the foot that we have associated with the ion Weibel instability. Using numerical experiments in which the ramp fluctuations are suppressed we have shown that this instability is associated with the beam of specularly reflected ions at their turnaround point at the upstream edge of the ion foot. We note that this modulation can be seen, albeit at low resolution, in the spectra for $B_x$ shown in \citet{lobu03} taken in the foot of a shock with $\theta_{Bn}=88^\circ$ and $M_\mathrm{A}=5.7$.

For 2-D simulations where the simulation plane is taken perpendicular to the coplanarity plane, field-parallel structure is suppressed, and instead the microstructure is dominated by fluctuations associated with a spatio-temporal pattern generated by a modulation of the reflected ion fraction \citep{busc07}. These fluctuations are not linked to any homogeneous plasma wave mode, and are characterized most strongly by the fact that they propagate in the direction and at the speed of the reflected ions at their upstream turnaround position. The relationship between these fluctuations and oscillations of the maximum of the shock overshoot and potential in 1-D \citep{leroal82}, and studies of shock reformation in 2-D \citep{yuanal09} has not yet been completely clarified.

In three-dimensional simulations there are signatures of several types of fluctuations seen in the 2-D simulations, indicating a complex, coupled system. There is a remnant of the zero-frequency modulation associated with the ion Weibel instability. However, the fluctuations are now dominated by a pattern of fluctuations generated by two counter-propagating waves inclined to the field direction. An analysis of the $\omega-k$-space representation shows that the fluctuations have propagation velocity components field-parallel (and anti-parallel) and, additionally, in the sense of motion of the reflected-gyrating ions. The inference is that fluctuations are carried by the reflected ions, and hence have a propagation velocity consistent with their cross-field motion. Thus the dominant microstructure seen in the 3-D simulations is the result of a coupling between the field-parallel fluctuations of the ``coplanarity plane'' 2-D simulations and the reflected-ion fluctuations of the ``perpendicular to coplanarity plane'' 2-D simulations. The identification of such fluctuations in spacecraft data will be challenging, but maybe possible with closely-spaced multi-spacecraft constellations, such as Cluster or MMS.

As discussed above, we have chosen shock parameters suited to study microstructure at the supercritical shock, as modelled by the hybrid plasma simulation method, in a relatively simple situation. Further simulations are required to study these processes in quasi-perpendicular geometries where there is no longer symmetry in the directions parallel and anti-parallel to the magnetic field. The effects of whistler structure generated at the shock ramp will also become more important as the shock normal angle is reduced. The role of nonlinear processes associated with shock reformation should be studied for the dependence on Mach number and ion beta, as these have been found to be crucial in 1-D hybrid simulations \citep{hellal02}. These issues will be addressed in future work.

This work was supported by the UK Science and Technology Facilities Council (STFC) grant ST/J001546/1. The research leading to the presented results has received funding from the European Commission's Seventh Framework Programme FP7 under the grant agreement SHOCK (project number 284515). This work used the DiRAC Complexity system, operated by the University of Leicester IT Services, which forms part of the STFC DiRAC HPC Facility (www.dirac.ac.uk). This equipment is funded by BIS National EInfrastructure capital grant ST/K000373/1 and STFC DiRAC Operations grant ST/K0003259/1. DiRAC is part of the National E-Infrastructure. This work was also supported by the projects RVO:67985815 and RVO:68378289 funded by the Grant Agency of the Czech Republic.

\appendix
\section{Ion Weibel instability}\label{appA}

There is an extensive literature for instabilities driven by cross-field drifts. \citet{chang:1990} considered the situation of a cold ion streaming perpendicular to a magnetic field, and, by including electromagnetic ion effects, found a purely growing (zero-frequency) mode for wavevectors parallel, or at small oblique angles, to the magnetic field. Their analysis assumed unmagnetized ions, so that an ion trajectory was treated as a straight line. One of the first applications of this instability was in the study of current disruption during substorm expansions in the geomagnetic tail \citep{lui:1991}. \citet{wu:1992} subsequently studied the instability more generally, and found that, even when relaxing the condition of unmagnetized ions and small $\beta_\mathrm{i}$, the zero-frequency instability persists, and in contrast to other cross-field drift instabilities (such as the modified two-stream instability) the growth rate remains high even for high $\beta_\mathrm{i}$. The mechanism of the instability is essentially that of the Weibel instability \citep{weibel:1959} as described by \citet{fried:1959} for counterstreaming electrons. For this reason \citet{chang:1990} dubbed it the ion Weibel instability (IWI). However, as \citet{wu:1992} note, for the case discussed here there is a net equilibrium current and the instability is due to a cross-field drift between electrons and ions, whereas the Weibel instability is usually driven by an electron temperature anisotropy in the low field limit.

A simple derivation of the growth rate can be obtained as follows. Consider the situation of a plasma consisting of a core population of ions and electrons, together with a cross-field ion beam. We take the magnetic field uniform in the $y$ direction $\boldsymbol{B} = B_0 \widehat{\boldsymbol{y}}$. At the front (upstream) edge of the shock foot, where the specularly reflected ions reach the turning point in their gyration, they form a single, cross-field beam which we take to have velocity $v_\mathrm{b} \widehat{\boldsymbol{z}}$ and density $n_\mathrm{b}$. The core ion density is $n_\mathrm{i}$ which is equal to that of the upstream flow; the electron density adjusts for charge neutrality. The density of reflected ions is typically of the order of 0.2-0.1$n_\mathrm{i}$. The foot is strongly inhomogeneous due to the flow deceleration, the presence of outgoing and returning reflected ions, and the increases of the magnetic field and shock potential. Nevertheless, at its upstream edge, the flow is not yet strongly disturbed, so we can use the upstream flow frame (where the core populations are stationary) for the instability analysis. We assume perturbations $\delta B_x$ and $\delta E_z$ (i.e., linearly polarized) which only have $y$ and $t$ dependence, so that they are coupled via the Faraday law $\partial_y \delta E_z = - \partial_t \delta B_x $. By linearizing the beam ion equation of motion, the conservation of number applied to the beam ions, and Ampere's law it is possible to derive a consistent set of equations. With the usual harmonic ansatz, the condition for non-trivial solutions is found to be $\omega^2 = - F(k)$ where the function $F(k)$ is positive for all $k$. This corresponds to a zero real frequency (i.e., non-propagating) mode with growth rate given by
\begin{equation}
\gamma = \left(\frac{n_\mathrm{b}}{n_\mathrm{i}} \right)^{1/2} \left( \frac{v_\mathrm{b}}{V_\mathrm{A}} \right)
 \left( \frac{(k\lambda_\mathrm{i})^2}{(k\lambda_\mathrm{i})^2 + n_\mathrm{b}/n_\mathrm{i}} \right)^{1/2}  \Omega_\mathrm{i} .
\end{equation}
Here $V_\mathrm{A}$ and $\lambda_\mathrm{i}$ are those of the upstream flow.

This simple derivation ignores the response from the core ion and electrons, so that the mode is carried entirely by the cross-field beam. Nevertheless, the dependence of the growth rate on $k$ is similar to that of the more complete analysis \citep[fig.~1,][]{chang:1990}. The growth rate $\gamma \rightarrow 0$ as $k\rightarrow 0$, and asymptotes as $k$ increases to 
\begin{equation}
\gamma \sim \left(\frac{n_\mathrm{b}}{n_\mathrm{i}} \right)^{1/2} \frac{v_\mathrm{b}}{V_\mathrm{A}} \Omega_\mathrm{i},
\end{equation}
which is similar to the expression from \citeauthor{chang:1990}, except for the factor depending on the beam density. (For the comparison, recall that $\omega_\mathrm{pi}/c = \Omega_\mathrm{i}/V_\mathrm{A}$.) For the straight line orbit approximation to be valid, $\gamma > \Omega_\mathrm{i}$, which for $n_\mathrm{b}/n_\mathrm{i} \sim 0.2$ and $v_\mathrm{b}/V_\mathrm{A}\sim 5$ is barely satisfied. However, as shown by \citet{wu:1992}, the instability persists even for magnetized ions. A typical scale for the instability can be estimated from the value $k_\mathrm{m}$ at which $\gamma$ achieves $1/\sqrt{2}$ of its asymptotic value, giving $k_\mathrm{m}\lambda_\mathrm{i} \sim 3.2 - 2.2$ for $n_\mathrm{b}/n_\mathrm{i} \sim 0.1 - 0.2$. This is approximately in the range found in the simulations.

This estimate of the characteristics of the instability depends on some drastic simplifications. The frame used for the analysis is the upstream flow frame, rather than the plasma frame which decelerates through the shock foot. Effects of inhomogeneity in the shock foot are neglected, and is also the fact that the reflected ions will have already had their orbits perturbed before they reach the upstream edge of the foot (at the turnaround point where their $x$ component of velocity is zero). Also, in the simulations there is evidence of oblique oriented phase fronts and coupling to the other magnetic field components, indicating that probably both inhomogeneity, growth of the IWI at oblique angles, and coupling to propagating oblique whistlers may play a role.

\bibliographystyle{jpp}

\bibliography{jppshock}

\end{document}